\documentclass[9pt,twocolumn,twoside]{osajnl}

\journal{josaa}
\setboolean{shortarticle}{false}
\usepackage{amsmath,widetext}
\usepackage{calligra}
\usepackage{color}

\title{Electromagnetic energy in multilayered spherical particles}

\author[1,*]{Ilia L. Rasskazov}
\author[2]{Alexander Moroz}
\author[1]{P. Scott Carney}

\affil[1]{The Institute of Optics, University of Rochester, Rochester, NY 14627, USA}
\affil[2]{Wave-scattering.com}

\affil[*]{Corresponding author: irasskaz@ur.rochester.edu}

\begin{abstract}
We obtain exact analytic expressions for (i) the electromagnetic energy radial density within and outside a multilayered sphere and (ii) the total electromagnetic energy stored within its core and each of its shells. Explicit expressions for the special cases of lossless core and shell are also provided. The general solution is based on compact recursive transfer-matrix method and its validity includes also magnetic media. The theory is illustrated on the examples of electric field enhancement within various metallo-dielectric silica-gold multilayered spheres. User-friendly MATLAB code which includes the theoretical treatment, is available as a supplement to the paper.
\end{abstract}

\setboolean{displaycopyright}{true}

\begin{document}

\maketitle

\section{Introduction}
Multilayered spherical particles of various sizes and material composition are an important part of modern science and technology due to exceptionally adjustable and extraordinary electromagnetic properties. In this regard, the interaction of the electromagnetic wave with a multilayered spherical particle under plane-wave~\cite{Aden1951,Toon1981,Wu1991,Bhandari1985,Mackowski1990,Sinzig1994,Bohren1998,Yang2003,Pena2009,Shore2015,Ladutenko2017} or general beam~\cite{Gouesbet1988,Onofri1995,Wu1997a,Li2007,Mojarad2009} illuminations represents a problem of a long-standing interest. The solution of this problem implies the definition of the electromagnetic field within or outside of a sphere which allows to get its absorption, scattering, extinction or other important characteristics. Among these properties, the cycle- and orientation-averaged electric $|{\bf E}|^2$ and magnetic $|{\bf H}|^2$ fields (in general, electromagnetic energy) within a particular layer (shell) or in the vicinity of a multilayered sphere is of great importance since it defines performance and suitability of a multilayered sphere for a large number of intriguing applications: nonlinear optics~\cite{Neeves1989,Pu2010,Butet2012}, lasing~\cite{Gordon2007,Noginov2009,Passarelli2016}, heating~\cite{Harris2006,Lukyanchuk2012,Neumann2013}, photocatalysis~\cite{Schlather2017}, fluorescence enhancement~\cite{Moroz2005,Moroz2005a,Ayala-Orozco2014a,Sakamoto2017}, plasmon-enhanced upconversion~\cite{Wu2014,Rasskazov18OMEx}, energy harvesting and storing~\cite{Moon2014,Meng2017,Phan2018a,Li2018c}, surface-enhanced Raman spectroscopy~\cite{Kodali2010a,Pena-Rodriguez2011,Khlebtsov2017}, biology and medicine~\cite{Jain2006,Khlebtsov2007,GhoshChaudhuri2012,Ayala-Orozco2014,Zakomirnyi17JQSRT}. 

Thus, a specific attention is drawn to theoretical considerations of the electromagnetic energy within and in a proximity of multilayered spherical particles. This fundamental problem has been thoroughly studied for homogeneous spheres: exact analytic expressions are reported for the electromagnetic energy in dielectric~\cite{Messinger1981,Bott1987}, magnetic~\cite{Arruda2010} and chiral~\cite{Arruda2013a} spheres. These solutions have been extended for two-layered~\cite{Gordon2007,Arruda2012} and three-layered~\cite{Khlebtsov2017} spheres by using the recursive relations, which makes corresponding analytic representation quite cumbersome and difficult to generalize for $N>3$ shells. Alternatively, semi-analytic approach might be used for estimating orientation-averaged local fields~\cite{Kodali2010a}, which involves \textit{analytic} representation of electromagnetic field within the spherical particle and its consequent \textit{numeric} integration within the volume of a particle~\cite{Peirce1957}. However, given that various optimization algorithms~\cite{Goldberg1989,Storn1997} are used to find the optimal design of multilayered sphere for a particular application, the development of closed-form analytic expressions becomes highly desirable. Here, we fulfill this need and present rigorous, and, quite importantly, compact analytic solution for (i) the electromagnetic energy radial density within and outside a multilayered sphere and (ii) the total electromagnetic energy within its core and each of its shells. 

The paper is organized as follows. In Sec.~\ref{sec:tmatrix}, we provide a brief overview of the recursive transfer-matrix solution of the electromagnetic light scattering from a multilayered sphere, which has been proposed and thoroughly discussed in Ref.~\cite{Moroz2005}; in Sec.~\ref{sec:energy}, within the framework of this formalism, we derive a solution for the electromagnetic energy and its density within the multilayered sphere; in Sec.~\ref{sec:energy_spec} we provide explicit expressions for specific cases of the electromagnetic energy stored within a core of multilayered sphere, in a lossless shell, and in surrounding medium close to a sphere. Discussion of numerical results for silica-gold multilayered nanospheres is given in Sec.~\ref{sec:disc}. Finally, we draw conclusive remarks in Sec.~\ref{sec:conc} and provide useful relations and derivations in the Appendix.

\section{Recursive Transfer-Matrix Method}
\label{sec:tmatrix}
\begin{figure}[b!]
    \centering
    \includegraphics{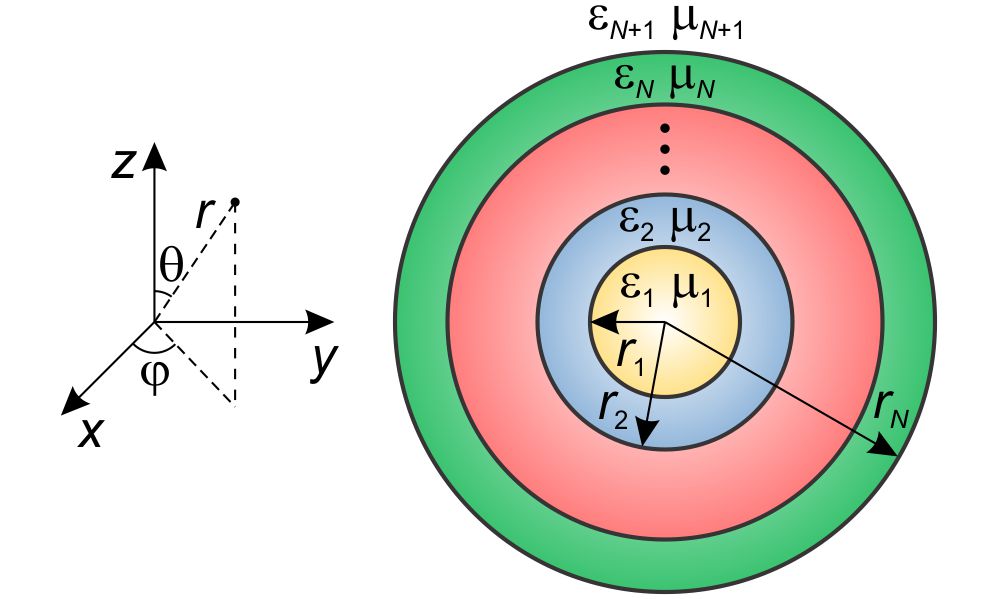}
    \caption{Schematic representation of the multilayered sphere embedded in homogeneous isotropic host medium with permittivity $\varepsilon_h=\varepsilon_{N+1}$ and permeability $\mu_h=\mu_{N+1}$.}
    \label{fig:scheme}
\end{figure}

Consider a multilayered sphere with $N$ concentric shells as shown in Fig.~\ref{fig:scheme}. The sphere core counts as shell with number $n=1$ and the host medium is the $n=N+1$ shell. Each shell is assumed to be homogeneous and isotropic with scalar permittivity $\varepsilon_n$ and permeability $\mu_n$. The outer radius of the $n$-th shell is denoted by $r_n$. Spherical coordinates are centered at the sphere origin.

We assume that the multilayered sphere is illuminated with a harmonic electomagnetic wave having 
vacuum wavelength $\lambda$. Corresponding wave vector in the $n$-th shell is $k_n = \eta_n\omega/c = 2\pi\eta_n/\lambda$, where $c$ is the speed of light in vacuum, $\omega$ is frequency, and $\eta_n = \sqrt{\varepsilon_n \mu_n}$ is the refractive index. Electromagnetic fields in any shell are described by the stationary macroscopic Maxwell's equations (in Gaussian units, with time dependence $e^{-i\omega t}$ assumed and suppressed throughout the paper):

\begin{equation}
{\bf E}=\frac{i c}{\omega\varepsilon}\,
         \left(\mbox{\boldmath $\nabla$}\times {\bf H}\right) \ , \qquad
{\bf H}= -\frac{i c}{\omega\mu}\,\left(\mbox{\boldmath $\nabla$}\times{\bf E}\right) \ ,
\label{mex}
\end{equation}
where the permittivity $\varepsilon$ and permeability $\mu$ are scalars.

Following the notation of \cite{Moroz2005}, the basis of normalized (the normalization here refers to angular integration) {\em transverse} vector multipole fields $\mbox{\boldmath $\nabla$}\cdot{\bf F}_{\gamma L}\equiv 0$ that satisfy the vector Helmholtz equation

\begin{equation}
\mbox{\boldmath $\nabla$}\times \left[\mbox{\boldmath $\nabla$} \times {\bf F}_{\gamma L}(k,{\bf r}) \right] = k^2 {\bf F}_{\gamma L}(k,{\bf r}) 
\nonumber
\end{equation}
for $n$-th shell, $1\leq n\leq N+1$, can be formed as~\cite{Moroz2005}:

\begin{equation}
\begin{split}
{\bf F}&_{ML}(k_n,{\bf r}) = f_{M L} (k_n r) {\bf Y}^{(m)}_L({\bf r}) \ , \\
{\bf F}&_{EL}(k_n,{\bf r}) = \\
& \frac{1}{k_n r}\left\{\sqrt{l(l+1)}f_{E L} (k_n r){\bf Y}^{(o)}_L({\bf r}) + \dfrac{\rm d}{{\rm d}r} \left[r f_{E L}(k_n r)\right]{\bf Y}^{(e)}_L({\bf r})\right\} \ ,
\end{split}
\label{fvmultip}
\end{equation}
where
\begin{equation}
\begin{split}
\tilde{\bf F}_{EL}(k_n,{\bf r}) & = \frac{1}{k_n}\mbox{\boldmath $\nabla$}
     \times {\bf F}_{ML}(k_n,{\bf r}) \ , \\
\tilde{\bf F}_{ML}(k_n,{\bf r}) & = \frac{1}{k_n}\mbox{\boldmath $\nabla$}
    \times {\bf F}_{EL}(k_n,{\bf r}) \ ,
\end{split}
\label{fvmultipr}
\end{equation}
with $\tilde{f}_{EL}=f_{ML}$ and $\tilde{f}_{ML}=f_{EL}$. Here $L=lm$ is a composite angular momentum index; ${\bf Y}^{(m)}_L$, ${\bf Y}^{(o)}_L$ and ${\bf Y}^{(e)}_L$ are magnetic, longitudinal, and electric vector spherical harmonics of degree $l$ and order $m$ (see Appendix \ref{sc:vsh} for their definition), and $f_{\gamma L}$ is a suitable linear combination of spherical Bessel functions. Provided that the multipole fields in \eqref{fvmultip} represent ${\bf E}$, the respective subscripts $M$ and $E$ denote the {\em magnetic}, or {\em transverse electric} (TE), and {\em electric}, or {\em transverse magnetic} (TM), polarizations~\cite{Jackson1999}. 

In the respective cases that $f_{\gamma l}=j_l$ and $f_{\gamma l}=h_l^{(1)}$, where $j_l$ and $h_l^{(1)}$ being the spherical Bessel functions of the first and third kind, correspondingly, the multipoles ${\bf F}_{\gamma L}$ will be denoted as ${\bf J}_{\gamma L}$ and ${\bf H}_{\gamma L}$. General solution for the electric field in the $n$-th shell, $1\leq n\leq N+1$, reads then as~\cite{Moroz2005}:
\begin{equation}
\begin{split}
{\bf E}_\gamma ({\bf r}) = & \sum_L  {\bf F}_{\gamma L}(k_n, {\bf r}) \\ 
= & \sum_{\gamma,L} \left[ A_{\gamma L} (n) {\bf J}_{\gamma L}(k_n, {\bf r}) + B_{\gamma L}(n) {\bf H}_{\gamma L}(k_n, {\bf r})\right] \ ,
\end{split}
\nonumber 
\end{equation}
with corresponding
\begin{equation}
f_{\gamma L}=A_{\gamma L} (n) j_l(k_n r)+ B_{\gamma L}(n) h^{(1)}_l(k_n r) 
\label{flcmb}
\end{equation} 
to be determined.
The expansion of magnetic field {\bf H} is related to that of the electric field {\bf E} by the stationary macroscopic Maxwell's equations (\ref{mex}) on using relations (\ref{fvmultipr}).

Expansion coefficients $A_{\gamma L}(n)$ and $B_{\gamma L}(n)$ are determined by matching fields across the shell interfaces, i.e. requiring that the tangential components of ${\bf E}$ and ${\bf H}$ are continuous. For surrounding medium, i.e. the $(N+1)$-th shell, the expansion coefficients will occasionally be written as

\begin{equation}
C_{\gamma L} \equiv A_{\gamma  L}(N+1) \ , \qquad D_{\gamma L} \equiv B_{\gamma  L}(N+1) \ .
\nonumber 
\end{equation}
Coefficients $A_{\gamma L}(n)$ and $B_{\gamma L}(n)$ can be found via transfer matrix solution in terms of $2\times 2$ {\em lowering} (backward) $T_{\gamma l}^-(n)$ and {\em raising} (forward) $T_{\gamma l}^+(n)$ transfer matrices~\cite{Moroz2005}. The {\em raising} transfer matrices translate the expansion coefficients $A_{\gamma  L}(n)$ and $B_{\gamma  L}(n)$ from the $n$-th shell into the coefficients $A_{\gamma  L}(n+1)$ and $B_{\gamma  L}(n+1)$ in the $(n+1)$-th shell:

\begin{equation}
\begin{pmatrix} A_{\gamma  L}(n+1)\\B_{\gamma  L}(n+1) \end{pmatrix} = T_{\gamma l}^+(n) \begin{pmatrix} A_{\gamma  L}(n)\\B_{\gamma  L}(n) \end{pmatrix} \ , 
\label{forwtrm}
\end{equation}
whereas the {\em lowering} transfer matrices translate the coefficients $A(n+1)$ and $B(n+1)$ in the $(n+1)$-th shell into the coefficients $A(n)$ and $B(n)$ in the $n$-th shell:

\begin{equation}
\begin{pmatrix} A_{\gamma  L}(n)\\B_{\gamma  L}(n) \end{pmatrix} = T_{\gamma l}^-(n) \begin{pmatrix} A_{\gamma  L}(n+1)\\B_{\gamma  L}(n+1) \end{pmatrix} \ .
\label{backtrm}
\end{equation}
It can be easily seen from \eqref{forwtrm} and \eqref{backtrm} that the {\em lowering} and {\em raising} transfer matrices are related as: 

\begin{equation}
\left[ T_{\gamma l}^+(n)\right]^{-1} = T_{\gamma l}^-(n) \ ,
\qquad
\left[ T_{\gamma l}^-(n)\right]^{-1} = T_{\gamma l}^+(n) \ .
\label{tinv}
\end{equation}
Provided that the coefficients $A_{\gamma L}(n+1)$ and $B_{\gamma L}(n+1)$ are known, the coefficients $A_{\gamma L}(n)$ and $B_{\gamma L}(n)$ can be unambiguously determined, and vice versa. The constituent transfer matrices $T_{\gamma l}^+$ and $T_{\gamma l}^-$ can be viewed as analagous to {\em ladder} operators of quantum mechanics.

To determine the expansion coefficients at any shell, {\em mixed} boundary conditions are imposed which fix {\em two} of the coefficients $A_{\gamma L}(n)$ and $B_{\gamma L}(n')$ for each given $\gamma$ and $L$, where in general $n\neq n'$:

\begin{enumerate}

\item The {\em regularity} condition of the solution at the sphere origin: 

\begin{equation}
B_{EL}(1)=B_{ML}(1)\equiv 0 \ ;
\label{dipregbc}
\end{equation}

\item For a given frequency $\omega$, the $A_{\gamma L}(N+1) = C_{\gamma L}$ coefficients are equal to the expansion coefficients of an incident electromagnetic field in spherical coordinates. 

\end{enumerate}

For general (e.g. focused or Gaussian) beams numerical integration is, as a rule,
required to arrive at the expansion coefficients~\cite{Gouesbet1988,Onofri1995,Li2007,Mojarad2009}.
Nevertheless, closed analytic expressions for the $C_{\gamma L}=0$'s are known for two important incident fields. The most familiar example, which we shall examine in detail below,
is furnished by incident plane 
electromagnetic wave ${\bf E}({\bf r}) = {\bf E}_0 \exp\left({i{\bf k}\cdot{\bf r}}\right)$, which expansion in vector spherical wave functions reads as 

\begin{equation}
{\bf E}_0 \exp\left(i{\bf k}\cdot{\bf r}\right) = \sum_L C_{\gamma L}\, {\bf J}_{\gamma L}(kr) \ ,
\label{pwe1}
\end{equation}
where

\begin{equation}
C_{ML} = 4\pi i^{l} {\bf E}_0 \cdot {\bf Y}^{(m)*}_L ({\bf k}) \ , \quad
C_{EL} = 4\pi i^{l-1} {\bf E}_0 \cdot {\bf Y}^{(e)*}_L ({\bf k}) \ ,
\label{pwcc}
\end{equation}
and asterisk denotes complex conjugate. In particular, for the plane wave incident along the $z$-axis and polarized along the $x$-axis (i.e. parallel to $\hat{\bf e}_\vartheta$ for $\varphi=0$):

\begin{equation}
\begin{split}
C_{ML} & = i^{l} \sqrt{(2l+1)\pi}\, E_0\, \delta_{m,\pm 1} \ , \\
C_{EL} & = \pm i^{l} \sqrt{(2l+1)\pi}\, E_0\, \delta_{m,\pm 1} \ ,
\nonumber 
\end{split}
\end{equation}
where $\delta_{mm'}$ is the Kronecker delta function. For an incident dipole field see Ref.~\cite{Moroz2005}. 

Irrespective of the incident field, one has the following explicit expressions for the constituent {\em backward} and {\em forward} transfer matrices~\cite{Moroz2005}:

\begin{widetext}
\begin{equation}
T^-_{Ml}(n) =
- i \begin{pmatrix}
  \tilde{\eta} \zeta_l'(x) \psi_l(\tilde{x}) - \tilde{\mu} \zeta_l(x)\psi_l'(\tilde{x}) 
& \tilde{\eta} \zeta_l'(x) \zeta_l(\tilde{x}) - \tilde{\mu} \zeta_l(x) \zeta_l'(\tilde{x}) \\
- \tilde{\eta} \psi_l'(x)\psi_l(\tilde{x}) + \tilde{\mu} \psi_l(x)\psi_l'(\tilde{x})
& - \tilde{\eta} \psi_l'(x) \zeta_l(\tilde{x}) + \tilde{\mu} \psi_l(x)\zeta_l'(\tilde{x})
\end{pmatrix} \ ,
\label{bacmtm}
\end{equation}

\begin{equation}
T^-_{El}(n) = 
- i \begin{pmatrix}
  \tilde{\mu} \zeta_l'(x)\psi_l(\tilde{x}) - \tilde{\eta} \zeta_l(x)\psi_l'(\tilde{x}) 
& \tilde{\mu} \zeta_l'(x) \zeta_l(\tilde{x}) - \tilde{\eta} \zeta_l(x) \zeta_l'(\tilde{x}) \\
- \tilde{\mu} \psi_l'(x)\psi_l(\tilde{x}) + \tilde{\eta} \psi_l(x)\psi_l'(\tilde{x})
& - \tilde{\mu} \psi_l'(x) \zeta_l(\tilde{x}) + \tilde{\eta} \psi_l(x)\zeta_l'(\tilde{x})
\end{pmatrix} \ ,
\label{bacetm}
\end{equation}

\begin{equation}
T^+_{Ml}(n) = 
- i \begin{pmatrix}
  \zeta_l'(\tilde{x}) \psi_l(x)/\tilde{\eta} - \zeta_l(\tilde{x})\psi_l'(x)/\tilde{\mu}
& \zeta_l'(\tilde{x}) \zeta_l(x)/\tilde{\eta} - \zeta_l(\tilde{x})\zeta_l'(x)/\tilde{\mu} \\
- \psi_l'(\tilde{x})\psi_l(x)/\tilde{\eta} + \psi_l(\tilde{x})\psi_l'(x) /\tilde{\mu}
& - \psi_l'(\tilde{x}) \zeta_l(x)/\tilde{\eta} + \psi_l(\tilde{x})\zeta_l'(x)/\tilde{\mu}
\end{pmatrix} \ ,
\label{formtm}
\end{equation}

\begin{equation}
T^+_{El}(n) =
- i \begin{pmatrix}
  \zeta_l'(\tilde{x})\psi_l(x)/\tilde{\mu}  - \zeta_l(\tilde{x})\psi_l'(x) /\tilde{\eta}
& \zeta_l'(\tilde{x}) \zeta_l(x)/\tilde{\mu} - \zeta_l(\tilde{x}) \zeta_l'(x) /\tilde{\eta} \\
- \psi_l'(\tilde{x})\psi_l(x)/\tilde{\mu} + \psi_l(\tilde{x})\psi_l'(x) /\tilde{\eta}
& - \psi_l'(\tilde{x}) \zeta_l(x)/\tilde{\mu} + \psi_l(\tilde{x})\zeta_l'(x) /\tilde{\eta}
\end{pmatrix} \ ,
\label{foretm}
\end{equation}
\end{widetext}
\noindent
where $\psi_l(x) = xj_l(x)$ and $\zeta_l(x) = x h^{(1)}_l(x)$ are the Riccati-Bessel functions, prime denotes the derivative with respect to the argument in parentheses, and 

\begin{equation}
x_n=k_n r_n , \quad \tilde{\eta}_n=\eta_n/\eta_{n+1} \ , \quad
\tilde{x}_n=x_n / \tilde{\eta}_n \ , \quad \tilde{\mu}_n = \mu_n/\mu_{n+1} \ . \nonumber
\end{equation}
For the sake of clarity, the $n$-subscript has been suppressed in \eqref{bacmtm}~--~\eqref{foretm}. The above relations for $T^-_{\gamma l}(n)$ and $T^+_{\gamma l}(n)$ are general and valid for any homogeneous and isotropic medium, including {\em magnetic} materials with $\mu_n \neq 1$.

It occurs that the formalism becomes compact if one introduces the composite transfer matrices ${\cal T}_{\gamma l} (n)$ and ${\cal M}_{\gamma l} (n)$ defined as ordered (from the left to the right) products of the constituent {\em forward} and {\em backward} $2\times 2$ matrices:

\begin{equation}
{\cal T}_{\gamma l} (n) = \prod_{j=n-1}^{1} T_{\gamma l}^+ (j) \ , \qquad {\cal M}_{\gamma l}  (n) = \prod_{j=n}^N T_{\gamma l}^- (j) \ .
\nonumber 
\end{equation}
Composite matrices ${\cal T}_{\gamma l} (n)$ and ${\cal M}_{\gamma l} (n)$ transfer expansion coefficients from the sphere core to the $n$-th shell, and from the surrounding medium to the $n$-th shell, respectively. Note that ${\cal T}_{\gamma l} (n)$ is defined for $2 \leq n\leq N+1$, while ${\cal M}_{\gamma l} (n)$ is defined for $1\leq n\leq N$. Analogously to \eqref{tinv}, the following relations can be applied:

\begin{equation}
\left[{\cal T}_{\gamma l} (N+1)\right]^{-1} = {\cal M}_{\gamma l}  (1) \ ,
\enspace
\left[{\cal M}_{\gamma l} (1)\right]^{-1} = {\cal T}_{\gamma l}  (N+1) \ .
\label{tminv}
\end{equation}

Note that the {\em regularity} condition given by \eqref{dipregbc}, unambiguously determines the $m$-independent ratio $D_{\gamma L}/C_{\gamma L}$~\cite{Jackson1999}:

\begin{equation}
D_{\gamma L}/C_{\gamma L} = {\cal T}_{21;\gamma l} (N+1)  /{\cal T}_{11;\gamma l} (N+1) \ . 
\label{dcrat} 
\end{equation}
Here ${\cal T}_{ij;\gamma l}(n)$ denotes the $(i,j)$-th element of the $2\times 2$ matrix ${\cal T}_{\gamma l}(n)$. 

Thus far, the electromagnetic field anywhere inside and outside a multilayered sphere is unambiguously determined from a pair of expansion coefficients $A(n)$ and $B(n)$ for the respective $n$-th shell (including the host medium denoted as $(N+1)$-th shell).

\section{Energy within a multilayered sphere}
\label{sec:energy}
The energy here will have the usual meaning of instant power integrated over a cycle of harmonic excitation \cite{Loudon1970,Ruppin1998,Ruppin2002,Nunes2011}. The \textit{total} electromagnetic energy $W$ within the multilayered sphere can be obtained by integrating the electromagnetic energy radial density $w_n(r)$ within each $n$-th shell and, consequently, summing up the total electromagnetic energies $W_n$ stored in each $n$-th shell:

\begin{equation}
   W = \sum_{n=1}^{N} W_n = \sum_{n=1}^{N} \int_{r_{n-1}}^{r_n} w_n(r) r^2 {\rm d} r \ ,
   \label{wdef}
\end{equation}
where $r_0=0$ for the core, and

\begin{equation}
   w_n(r) = \dfrac{1}{4} \oint \left[ G_e\left(\varepsilon_n\right) \left| {\bf E}({\bf r}) \right|^2 + G_m\left(\mu_n\right)\left| {\bf H} ({\bf r})\right|^2 \right] {\rm d} \Omega \ .
   \label{wdendef}
\end{equation}
Here $G_e\left(\varepsilon_n\right) = {\rm Re}\left(\varepsilon_n\right)$ and $G_m\left(\mu_n\right) = {\rm Re}\left(\mu_n\right)$ in the non-dispersive case~\cite{Bott1987}. For dispersive and absorbing  metallic shells $G_m$ remains the same, while $G_e$ can be described by Loudon's formula~\cite{Loudon1970}: $G_e=[{\rm Re}(\varepsilon_n) + 2\omega {\rm Im} (\varepsilon_n)/\Gamma_n]$, with $\Gamma_n$ being the free electron damping constant in the Drude formula. 

Angular integration of $\left| {\bf E} \right|^2$ and $\left| {\bf H} \right|^2$ in \eqref{wdendef} in any shell can be performed as follows (see Appendix \ref{sc:angint} for details): 

\begin{equation}
\begin{split}
\oint \left| {\bf E} \right|^2 {\rm d} \Omega = \sum_{l=1}^{\infty} \sum_{m=-l}^l & \left[ |f_{Mlm}|^2 + \frac{l+1}{2l+1} |f_{E,l-1,m}|^2 \right. \\
& \left. + \frac{l}{2l+1} |f_{E,l+1,m}|^2\right] \ ,
\end{split}
\label{e2}
\end{equation}
\begin{equation}
\begin{split}
\oint \left| {\bf H} \right|^2 {\rm d} \Omega = \frac{|\varepsilon_n|}{|\mu_n|} \sum_{l=1}^{\infty} \sum_{m=-l}^l & \left[|f_{Elm}|^2 + \frac{l+1}{2l+1} |f_{M,l-1,m}|^2 \right. \\ & \left. + \frac{l}{2l+1} |f_{M,l+1,m}|^2\right] \ .
\end{split}
\label{h2}
\end{equation}
The spherical Bessel functions of the order $l\pm 1$ here originate from eliminating the radial derivation in ${\bf F}_{EL}$ of Eq. (\ref{fvmultip}) by the identity (\ref{bbff}) of Appendix \ref{sc:angint}.

Below we present exact analytic expressions for each of the above quantities, i.e. $w_n(r)$, $W_n$, and $W$. There are two important steps. The first is the summation over $m$ in \eqref{e2} and \eqref{h2}, which we perform by (i) factorizing $f_{\gamma L}$, and then (ii) using sum rules for $C_{\gamma L}$. The second is a radial integration in~\eqref{wdef}, which we implement by using Lommel's integration formulas of Appendix \ref{sc:lomm}.

\subsection{Factorization of expansion coefficients}
Expansion coefficients $A_{\gamma L}(n)$ and $B_{\gamma L}(n)$ can be reformulated via forward ${\cal T}_{\gamma l}(n)$ and backward ${\cal M}_{\gamma l}(n)$ composite transfer matrices as follows:

\begin{equation}
\begin{split}
A_{\gamma L}(n) = & {\cal M}_{11;\gamma l} (n) C_{\gamma L} + {\cal M}_{12;\gamma l} (n) D_{\gamma L} \\
= & C_{\gamma L} \left[{\cal M}_{11;\gamma l} (n) + {\cal M}_{12;\gamma l} (n) \dfrac{{\cal T}_{21;\gamma l} (N+1)}{{\cal T}_{11;\gamma l} (N+1)} \right] \ ,
\end{split}
\label{acreln}
\end{equation}

\begin{equation}
\begin{split}
B_{\gamma L}(n) = & {\cal M}_{21;\gamma l} (n) C_{\gamma L} + {\cal M}_{22;\gamma l} (n) D_{\gamma L} \\
= & C_{\gamma L} \left[{\cal M}_{21;\gamma l} (n) + {\cal M}_{22;\gamma l} (n) \dfrac{{\cal T}_{21;\gamma l} (N+1)}{{\cal T}_{11;\gamma l} (N+1)} \right] \ .
\end{split}
\label{bcreln}
\end{equation}
Note that in this representation, coefficients $A_{\gamma L}$ and $B_{\gamma L}$ are products of $m$-dependent $C_{\gamma L}$ and $m$-independent expression in square brackets. Due to the linearity of the equations and spherical symmetry of the problem (the latter being reflected in $m$-{\em independent} transfer matrix elements), each of the expansion coefficients $A_{\gamma L}$ and $B_{\gamma L}$ can be factorized as a product of a $m$-{\em dependent} factor resulting from $m$-dependence of the expansion coefficients $C_{\gamma L}$, and a $m$-{\em independent} factor coming from the transfer matrices:

\begin{equation}
    A_{\gamma L}(n) = C_{\gamma L} \bar{A}_{\gamma l}(n) \ , \quad
    B_{\gamma L}(n) = C_{\gamma L} \bar{B}_{\gamma l}(n) \ .
\nonumber
\end{equation}
Hence each $f_{\gamma L}$ factorizes as

\begin{equation}
f_{\gamma L} = C_{\gamma L} \bar{f}_{\gamma l} \ ,
\label{fcte}    
\end{equation}
where $m$-{\em independent} $\bar{f}_{\gamma l}$ is explicitly defined as

\begin{equation}
    \begin{split}
    \bar{f}_{\gamma l} = \bar{A}_{\gamma l}(n) j_l(k_n r) + \bar{B}_{\gamma l}(n) h^{(1)}_l(k_n r) \ .
    \end{split}
\label{flcmbr}
\end{equation}
Note that in \eqref{e2} and \eqref{h2} one has to consider the radial solutions as indexed by $\gamma$, $l$, and $l \pm 1$, 

\begin{equation}
\bar{f}_{\gamma l\pm 1}(n) = \bar{A}_{\gamma l} (n) 
j_{l\pm 1}(k_n r)
         + \bar{B}_{\gamma l}(n) h^{(1)}_{l\pm 1} (k_n r) \ .
\label{flcmbrpm}
\end{equation}
Since the spherical Bessel functions of the order $l\pm 1$ in \eqref{e2} and \eqref{h2} originate from eliminating the radial derivation in ${\bf F}_{EL}$ of \eqref{fvmultip} according to the identity (\ref{bbff}) of Appendix \ref{sc:angint}, they are multiplied by the expansions coefficients $\bar{A}_{\gamma l}$ and $\bar{B}_{\gamma l}$ carrying the index $l$.

After the factorization given by \eqref{fcte}, integral in \eqref{e2} reads as:

\begin{equation}
\begin{split}
\oint & \left| {\bf E} \right|^2 \, {\rm d} \Omega = \sum_{l=1} \left[  |\bar{f}_{Ml}|^2  \sum_{m=-l}^l |C_{M lm}|^2 \right. \\
& + \left. \left(\frac{l+1}{2l+1}\, |\bar{f}_{E,l-1}|^2  + \frac{l}{2l+1}\, |\bar{f}_{E,l+1}|^2\right)\sum_{m=-l}^l |C_{E lm}|^2 \right] \ .
\end{split}
\label{e2p}
\end{equation}
Note that \eqref{h2} is factorized in a similar manner.

\subsection{Sum rules}
\label{sc:sumr}
One can eliminate the $m$-dependence in \eqref{e2p} recalling \eqref{pwcc} for $C_{\gamma l m}$ and using the sum rules~\cite{Mishchenko1991,Pendleton2001} for magnetic ${\bf Y}^{(m)}_L({\rm r})$ and electric ${\bf Y}^{(e)}_L({\rm r})$ vector spherical harmonics:

\begin{equation}
    \sum_{m=-l}^{l} {\bf Y}_L({\rm r}) \otimes {\bf Y}^{*}_L({\rm r}) = \dfrac{2l+1}{8\pi} \left( \hat{\bf e}_\vartheta \otimes \hat{\bf e}_\vartheta + \hat{\bf e}_\varphi \otimes \hat{\bf e}_\varphi \right) \ ,
\nonumber
\end{equation}
where $\otimes$ denotes the tensor product. Thus for plane wave incidence the $m$-dependence in \eqref{e2p} yields

\begin{equation}
\sum_{m=-l}^l  |{\bf E_0}\cdot {\bf Y}^{(m,e)*}_{L} |^2 = \frac{2l+1}{8\pi }  \left( |{\bf E}_\theta |^2 + |{\bf E}_\phi |^2  \right) = \frac{2l+1}{8\pi } |E_0|^2 \ ,
\nonumber 
\end{equation}
which is also applicable to a factorized representation of \eqref{h2}.
For an incident dipole field see Ref. \cite{Moroz2005}.

\subsection{Electromagnetic energy radial density}
After the factorization of \eqref{e2} and \eqref{h2}, and subsequent summation over the magnetic number $m$, we end up with the following expressions for the electric

\begin{equation}
\begin{split}
& \oint \left| {\bf E} \right|^2 \, {\rm d} \Omega = 2 \pi |E_0|^2 \\
& \times \sum_{l=1}^\infty \left[(2l+1)\, |\bar{f}_{Ml}|^2 + (l+1)\, |\bar{f}_{E,l-1}|^2  + l\, |\bar{f}_{E,l+1}|^2\right]
\end{split}
\label{e2m}
\end{equation}
and for magnetic

\begin{equation}
\begin{split}
& \oint \left| {\bf H} \right|^2 \, {\rm d} \Omega = 2 \pi |E_0|^2  \frac{|\varepsilon_n|}{|\mu_n|} \\
& \times \sum_{l=1}^\infty \left[ (2l+1)\, |\bar{f}_{El}|^2 + (l+1)\, | \bar{f}_{M,l-1}|^2  + l\, |\bar{f}_{M,l+1}|^2 \right]
\end{split}
\label{h2m}
\end{equation}
components of the electromagnetic field.

Thus, the electromagnetic energy radial density in~\eqref{wdendef} is explicitly defined with analytic expressions in \eqref{e2m} and \eqref{h2m}.

\subsection{Total electromagnetic energy}
Finally, the radial integration of \eqref{e2m} and \eqref{h2m} in \eqref{wdef} can be performed by using Lommel's integration formulas (see Appendix \ref{sc:lomm} for details):

\begin{equation}
\begin{split}
& \int_{r_{n-1}}^{r_n} r^2 {\rm d}r \oint \left| {\bf E} \right|^2 {\rm d}\Omega =
2 \pi |E_0|^2 \frac{r^3}{x^2-x^{*2}} \\
&\left. \times \sum_{l=1}^\infty \left[ 
(2l+1) \bar{F}_{Ml} + (l+1) \bar{F}_{E,l-1} + l \bar{F}_{E,l+1}
\right]\right|_{r=r_{n-1}}^{r=r_n} \ ,
\end{split}
\label{e2mi}
\end{equation}

\begin{equation}
\begin{split}
& \int_{r_{n-1}}^{r_n} r^2 {\rm d} r \oint \left| {\bf H} \right|^2  {\rm d}\Omega = 
2 \pi |E_0|^2  \frac{r^3}{x^2-x^{*2}} \frac{|\varepsilon_n|}{|\mu_n|} \\
& \left. \times \sum_{l=1}^\infty \left[ 
(2l+1) \bar{F}_{El} + (l+1) \bar{F}_{M,l-1}  + l \bar{F}_{M,l+1}
\right]\right|_{r=r_{n-1}}^{r=r_n} \ .
\end{split}
\label{h2mi}
\end{equation}
Here $x = k_n r$, and purely {\em imaginary} functions 

\begin{equation}
\begin{split}
\bar{F}_{\gamma l} & = x \bar{f}_{\gamma,l+1}(x) \bar{f}^*_{\gamma l}(x) - x^* \bar{f}_{\gamma l}(x) \bar{f}^*_{\gamma,l+1}(x) \\
& = 2i \mbox{Im}\left[x \bar{f}_{\gamma,l+1}(x) \bar{f}^*_{\gamma l}(x) \right]
\end{split}
\label{fpl}
\end{equation}
are cancelled by purely \textit{imaginary} $x^2-x^{*2}=4i {\rm Re}(x) {\rm Im}(x)$ in the denominator, which results in purely \textit{real} integrals in \eqref{e2mi} and \eqref{h2mi}.

Substitution of \eqref{e2mi} and \eqref{h2mi} into \eqref{wdef} and \eqref{wdendef} yields 
in explicit expression for the total electromagnetic energy $W_n$ stored within each shell of 
the multilayered sphere. The above relations are general and valid within any shell including $(N+1)$-th layer being a surrounding medium.

\subsection{Normalized electromagnetic energy}
In some cases, it is of practical use to estimate \textit{normalized} electromagnetic energy instead of its absolute value. For example, the electromagnetic energy enhancement determines the performance of the spherical particle in surface-enhanced Raman spectroscopy~\cite{Kodali2010a,Khlebtsov2017}, plasmon-enhanced upconversion~\cite{Rasskazov18OMEx}, or might be important in other cases~\cite{Xu2005,Miroshnichenko2010a,Arruda2013b}. To get the corresponding enhancement factor, one could compare the energy stored within the $n$-th layer compared to the energy stored in a lossless host medium of the same volume. Given that in a homogeneous medium $|{\bf H}|^2 = (|\varepsilon_h|/|\mu_h|)|{\bf E}|^2$, the angularly integrated electromagnetic energy density (\ref{wdendef}) of the incident wave reduces in the {\em lossless} host medium to

\begin{equation}
w_0 = w_0^{(e)} + w_0^{(m)}= 2 \pi |E_0|^2 \varepsilon_h \ ,
\quad w_0^{(e,m)} = \pi |E_0|^2 \varepsilon_h \ ,
\nonumber 
\end{equation}
where $w_0^{(e)}$ and $w_0^{(m)}$ represent electric and magnetic components of the $w_0$, respectively. These quantities can be used to normalize the electromagnetic energy radial density $w_n(r)$ in the presence of a general multilayered sphere at the radial distance $r$ from the sphere origin. 

On using~\eqref{wdef}, the {\em total} electromagnetic energy stored within the shell with thickness $\left(r_n - r_{n-1}\right)$ characterized by a {\em lossless} $\varepsilon_h$, is defined as:

\begin{equation}
W_{0n} = \frac{2}{3} \pi |E_0|^2 \left(r_n^3 - r_{n-1}^{3}\right) \varepsilon_h \ .
\nonumber 
\end{equation}
This quantity can be used to normalize the {\em total} electromagnetic energy $W_n$ stored within $n$-th shell.

\subsection{Convergence criterion}
For the completeness of the developed theory, it is insightful to provide general remarks on the summation over $l$ in Eqs.~(\ref{e2m})--(\ref{h2mi}). Numerical implementation of these equations requires the truncation to some finite number $l_{\rm max}$, which can be defined for a particular value of the size parameter $x=kr$ with a widely used Wiscombe criterion~\cite{Wiscombe1980}:

\begin{equation}
l_{\rm max} =
\begin{cases}
x + 4x^{1/3} + 1 \ , \qquad 0.02 \leq x \leq 8 \\
x + 4.05x^{1/3} + 2 \ , \qquad 8 < x < 4200 \\
x + 4x^{1/3} +2 \ , \qquad 4200<x<20,000 \ .
\end{cases}    
\nonumber
\end{equation}
However, this criterion may vary for near- and far-fields~\cite{Allardice2014}. 

For large values of the size parameter, one could face convergence issues, because the theoretical treatment for multilayered spherical particles inevitably involves calculation of the difference of the products of the Riccati-Bessel functions. The most successful way to mitigate these issues is to factorize Riccati-Bessel functions with their logarithmic derivatives as shown in Ref.~\cite{Yang2003}. 

\section{Special Cases}
\label{sec:energy_spec}
Although presented formalism is rigorous and valid for a general multilayered sphere, it is insightful to discuss some special cases and provide corresponding explicit expressions, which might be more handy to use.

\subsection{Core region}
For core region, with $n=1$, \eqref{acreln} reduces on using \eqref{tminv} to

\begin{equation}
A_{\gamma L}(1) = \frac{C_{\gamma L} }{{\cal T}_{11;\gamma l} (N+1)} \ .
\label{acrel}
\end{equation}
Note that $B_{\gamma L} = 0$ due to regularity condition given in \eqref{dipregbc}. Thus, on using \eqref{acrel} and Eqs.~(\ref{fcte})--(\ref{flcmbrpm}):

\begin{equation}
\bar{f}_{\gamma l} = \frac{j_l (k_1r) }{ {\cal T}_{11;\gamma l} (N+1)} \ , \qquad
\bar{f}_{\gamma l\pm 1} = \frac{j_{l\pm 1} (k_1r) }{ {\cal T}_{11;\gamma l} (N+1)} \ .
\label{fbar}
\end{equation}
We emphasize that $\bar{f}_{\gamma l}$ and $\bar{f}_{\gamma l\pm 1}$ have the same denominator according to \eqref{flcmbr} and \eqref{flcmbrpm}. 

Thus, the {\em total} electromagnetic energy stored inside a core of a general multilayered sphere extending from $r=0$ to $r=r_1$ reads as:

\begin{equation}
\begin{split}
& W_1 = \int_0^{r_1} w_1(r) r^2 {\rm d}r = 
\frac{\pi |E_0|^2}{2} \frac{r_1^3}{x_1^2-x_{1}^{*2}} \\
& \times \sum_{l=1}^\infty
\left\{ (2l+1) \bar{F}^{(1)}_{l}
\left[ \frac{G_e\left(\varepsilon_1\right)}{|{\cal T}_{11;M l} (N+1)|^2} +
\frac{|\varepsilon_1|}{|\mu_1|} \frac{G_m\left(\mu_1\right)}{ |{\cal T}_{11;El} (N+1)|^2}
\right]
\right. \\
& \left. +
 (l+1) \bar{F}^{(1)}_{l-1}
\left[ \frac{G_e\left(\varepsilon_1\right)}{|{\cal T}_{11;El} (N+1)|^2} +
\frac{|\varepsilon_1|}{|\mu_1|} \frac{G_m\left(\mu_1\right)}{|{\cal T}_{11;Ml} (N+1)|^2}
\right] \right. \\
& \left. + 
l \bar{F}^{(1)}_{l+1}
\left[ \frac{G_e\left(\varepsilon_1\right)}{ |{\cal T}_{11;El} (N+1)|^2} +
\frac{|\varepsilon_1|}{|\mu_1|} \frac{G_m\left(\mu_1\right)}{|{\cal T}_{11;Ml} (N+1)|^2}
\right]
\right\} \ ,
\end{split}
\label{wft}
\end{equation}
where $\bar{F}^{(1)}_{l} = 2i{\rm Im}\left[ x_1 j_{l+1}(x_1) j_l^*(x_1) \right]$. 

Known results for the electromagnetic energy within a non-magnetic~\cite{Bott1987} and magnetic~\cite{Arruda2010} homogeneous sphere can be recovered from \eqref{wft} by considering the special case of $N=1$. 

\subsection{Lossless shell}
The lossless shell case can be obtained by taking the limit ${\rm Im}(\eta) \to 0$, which yields in $x=x^*$, and, as a consequence, vanishing of the denominator in \eqref{e2mi} and \eqref{h2mi}. After applying l'H\^{o}pital's rule (see Appendix \ref{sc:lomm} for details), \eqref{e2mi} and \eqref{h2mi} read as: 

\begin{equation}
\begin{split}
& \int_{r_{n-1}}^{r_n} r^2 {\rm d}r \oint \left| {\bf E} \right|^2 {\rm d}\Omega 
   =
 \pi |E_0|^2 \frac{r^3}{x} \\
& \times \left. \sum_{l=1}^\infty
          \left[ 
(2l+1) \bar{\Lambda}_{Ml} +
                 (l+1) \bar{\Lambda}_{E,l-1} + l \bar{\Lambda}_{E,l+1}
\right]\right|_{r=r_{n-1}}^{r=r_n} \ ,
\end{split}
\label{e2mif}
\end{equation}

\begin{equation}
\begin{split}
& \int_{r_{n-1}}^{r_n} r^2 {\rm d}r \oint \left| {\bf H} \right|^2 {\rm d}\Omega 
   =
\pi |E_0|^2 \frac{r^3}{x} 
 \frac{|\varepsilon_n|}{|\mu_n|} \\
& \times \left. \sum_{l=1}^\infty 
\left[ 
(2l+1) \bar{\Lambda}_{El} +
         (l+1) \bar{\Lambda}_{M,l-1}  + l \bar{\Lambda}_{M,l+1}
\right]\right|_{r=r_{n-1}}^{r=r_n} \ ,
\end{split}
\label{h2mif}
\end{equation}
where $r_0=0$, and $m$-independent parameter

\begin{equation}
\bar{\Lambda}_{\gamma l} = x \left(|\bar{f}_{\gamma l}|^2 + |\bar{f}_{\gamma l+1}|^2\right) 
         - (2l+1) \mbox{Re}\left( \bar{f}_{\gamma l} \bar{f}_{\gamma l+1}^*\right) \ .
\label{ldsh}
\end{equation}
Of note, Eqs.(\ref{e2mif})--(\ref{ldsh}) are also valid for a lossless core, though, appropriate expressions for $\bar{f}_{\gamma l}$ and $\bar{f}_{\gamma l+1}$ from \eqref{fbar} have to be used.

\subsection{Electromagnetic field in the vicinity of a sphere}
Finally, for many applications it is of interest to get the angular averaged electric or magnetic field intensity outside a multilayered particle. Assuming the plane wave incidence and the corresponding plane wave expansion in vector spherical wave functions given by \eqref{pwe1}, the electric field outside a spherical particle is defined as: 

\begin{equation}
\begin{split}
{\bf E} ({\bf r}) = & \sum_{\gamma,L} \left[
C_{\gamma L}  {\bf J}_{\gamma L}(k_h, {\bf r}) + D_{\gamma L}
{\bf H}_{\gamma L}(k_h, {\bf r})\right] 
= \\
& \sum_L {\bf F}_{\gamma L}(k_n, {\bf r})
= 
\sum_L C_{\gamma L}  \bar{{\bf F}}_{\gamma L}(k_h, {\bf r}) \ ,
\end{split}
\label{genesolo}
\end{equation}
where $f_{\gamma L}$ in ${\bf F}_{\gamma L}$ are given by~\cite{Moroz2005}:

\begin{equation}
\begin{split}
f_{\gamma L} & = C_{\gamma L} j_l(k_n r)+ D_{\gamma L} h^{(1)}_l(k_n r) \\
& = C_{\gamma L} \left[ j_l(k_h r) + h_l(k_h r) \dfrac{{\cal T}_{21;\gamma l} (N+1)}{{\cal T}_{11;\gamma l} (N+1)}\right] = C_{\gamma L} \bar{f}_{\gamma l} \ .
\end{split}
\label{fo}
\end{equation}
Here the $m$-{\em dependent} coefficients $C_{\gamma L}$ are given by \eqref{pwcc} and \eqref{dcrat} for $m$-{\em independent} ratio $D_{\gamma L}/C_{\gamma L}$ has been used. 

One can now proceed as before when arriving at \eqref{e2m}, to yield formally in the same formula, but with $\bar{f}_{\gamma l}$ given in \eqref{fo}. Corresponding enhancement of the average electric field intensity at the distance $r>r_N$ from the center of a spherical particle can be obtained after normalization \eqref{e2m} to $4\pi |E_0|^2$. The magnetic field enhancement near a multilayered sphere can be obtained analogously.

\section{Discussion}
\label{sec:disc}
For the sake of illustration, we apply theory developed here to metallo-dielectric multilayered nanospheres. Appropriately designed by well-developed fabrication procedures~\cite{Zhou1994,Graf2002}, such nanoparticles attract significant attention since they may extraordinarily absorb~\cite{Hasegawa2006,Grigoriev2015,Ladutenko2015}, scatter~\cite{Ruan2011} or transmit~\cite{Rohde2007,Alu2008,Hudak2019} electromagnetic irradiation, and serve as a platform for photonic bandgap structures~\cite{Moroz1999,Moroz2000,Moroz2002,Smith2002} or hyperbolic media~\cite{Wang2018h}.

\begin{figure}[t!]
    \centering
    \includegraphics{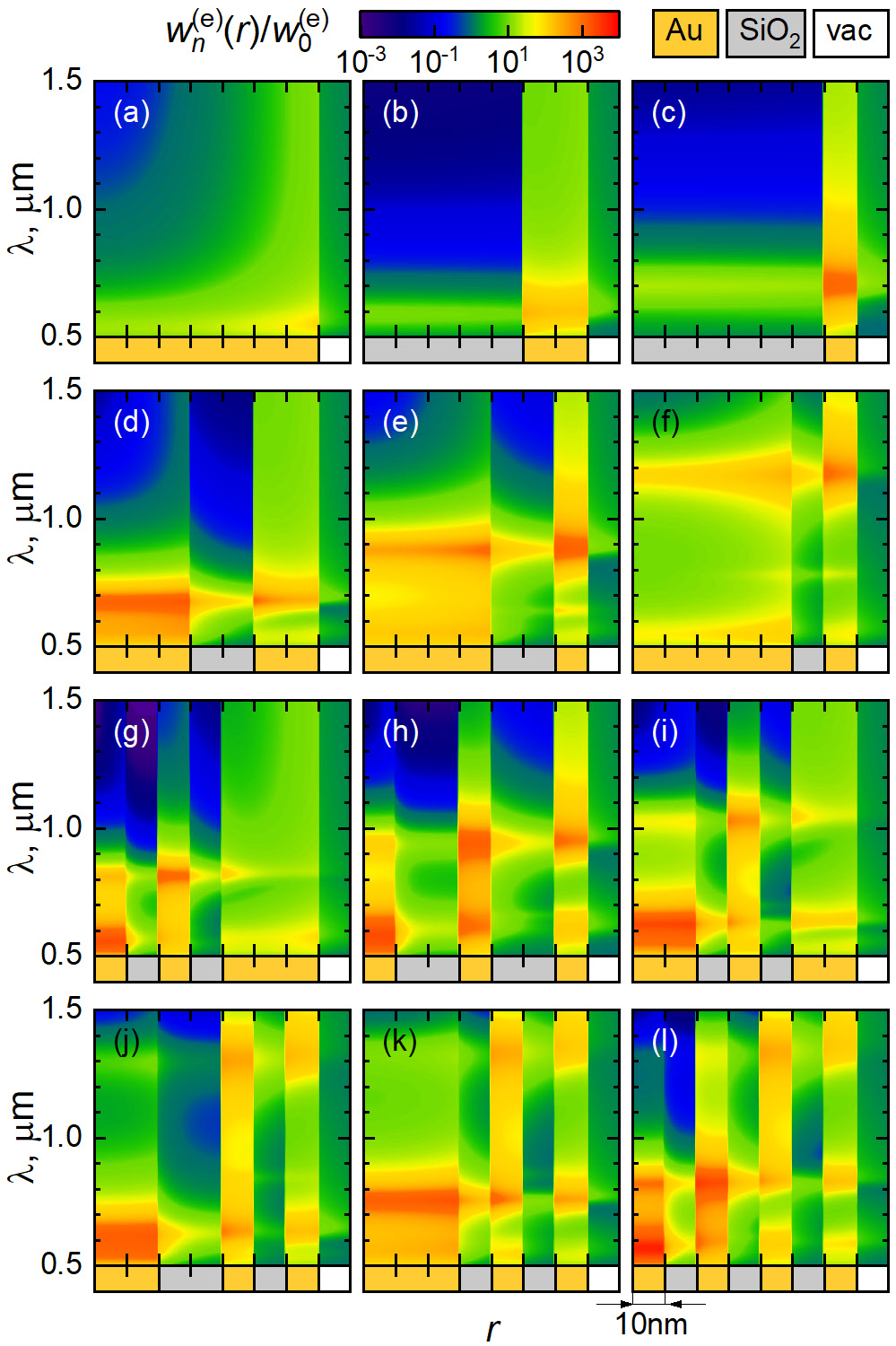}
    \caption{Normalized \textit{electric} energy radial density within and in the vicinity of the multilayered ${\rm SiO}_2$-Au spherical particles of fixed radius $r_N=70$~nm embedded in a vacuum with $\varepsilon_h=\mu_h=1$. Refractive index of ${\rm SiO}_2$ is assumed to be 1.45, while the experimental data from Ref.~\cite{Johnson1972} has been used for the refractive index of Au, with the electron mean free path correction~\cite{Moroz2008,Ruppin2015} for thin shells. Loudon's formula~\cite{Loudon1970} for $G_e$ has been employed with corresponding data for Drude model of Au~\cite{Blaber2009}. Different spheres composition, as denoted with colored bars on the bottom of each plot, with the rightmost 10~nm corresponding to the surrounding vacuum, have been considered: (a)~solid Au sphere, (b) and (c)~${\rm SiO}_2$/Au, (d)-(f)~Au/${\rm SiO}_2$/Au, (g)-(k)~Au/${\rm SiO}_2$/Au/${\rm SiO}_2$/Au, and (l)~Au/${\rm SiO}_2$/Au/${\rm SiO}_2$/Au/${\rm SiO}_2$/Au multilayered spheres.}
    \label{fig:ausio2}
\end{figure}

Figure~\ref{fig:ausio2} shows the normalized electric energy radial density $w_{n}^{(e)}(r)/w^{(e)}_0$ (which corresponds to the normalized electric term in \eqref{wdendef}) for widely used silica-gold nanospheres. $w_{n}^{(e)}(r)/w^{(e)}_0$ is shown for a number of ${\rm SiO}_2$-Au spheres of different composition, consisting of $n=1,2,3,5$ or 7 layers with varying thickness. It can be seen that our theory makes it convenient to investigate the electromagnetic field localization within the multilayered spheres. In a particular case of our study, one can observe wavelength-dependent features of the electric field localization within the various layers, depending on the particle composition: $w_{n}^{(e)}(r)/w^{(e)}_0$ acquires maximum values at different $\lambda$, from visible to near-IR, and within different layers. Taking as a benchmark the immediate exterior of a homogeneous Au sphere, significant energy density enhancements can be observed  both inside and outside of multilayered particles.

\section{Conclusion}
\label{sec:conc}
We have presented a self-consistent rigorous theory for the electromagnetic energy within a general multilayered sphere, which is applicable to a general illumination. Our main focus was on a plane wave illumination for which we obtained exact analytic expressions for (i) the  electromagnetic energy radial density within and outside a multilayered sphere and (ii) the total electromagnetic energy stored within its core and each of its shells. Other types of excitation~\cite{Moroz2005,Gouesbet1988,Onofri1995,Li2007,Mojarad2009,Moreira2016} require substitution of corresponding expressions of expansion coefficients $C_{\gamma L}$ for those in \eqref{pwcc}.

The reported formalism is valid for a wide range of sphere sizes and materials, including magnetic materials. Multilayered spheres from anisotropic~\cite{You-LinGeng2005,Qiu2007} or chiral~\cite{Arruda2013a} materials can be also considered with the presented formalism after modification.

The theory developed here could have numerous applications. The most straightforward are heating and nonlinear optics applications, which require the determination of $|{\bf E}|^2$ or its higher powers. Although only the electric part of the electromagnetic energy has been traditionally extensively considered in the literature, the recent development of all-dielectric nanophotonics also paves a way to a variety of exciting phenomena based on the manipulation of $|{\bf H}|^2$~\cite{Baranov2017a,Li2017e,Wiecha2019}, which also can be realized with the multilayered spherical particles of the appropriate composition~\cite{Tzarouchis2018}. In any case, a proper understanding of energy density distribution may provide a valuable insight in many other situations, some of which are the subject of future investigation.

Corresponding MATLAB code, which includes the theoretical treatment reported in this paper, is freely available on GitLab~\cite{stratify}.
It is straightforward to modify the code to obtain separate contributions for field intensities, and to determine the {\em stored} and {\em dissipated} energies for dispersive and absorbing media~\cite{Loudon1970,Ruppin1998,Ruppin2002,Nunes2011} (cf. also Appendix C of Ref.~\cite{Moroz2005}). It is worthwhile to emphasize that a numerical implementation of the developed theory is of low computational cost due to the utilization of explicit analytic expressions for the electromagnetic energy density.

\section*{Appendix}

\subsection{Vector spherical harmonics}
\label{sc:vsh}
The definition of the vector spherical harmonics from \eqref{fvmultip} varies in the literature (compare, for instance Bohren~\cite{Bohren1998}, Jackson~\cite{Jackson1999}, Chew~\cite{Kerker1980}, Mishchenko~\cite{Mishchenko2002}). Here we provide a ``combined'' representation for magnetic, longitudinal, and electric vector spherical harmonics  in \textit{spherical} $(\varphi, \vartheta, r)$ coordinates:

\begin{align}
{\bf Y}^{(m)}_L &= i \sqrt{ \frac{(l-m)!}{(l+m)!} } \sqrt{\frac{2l+1}{4\pi l(l+1)}} \nonumber \\
& \times \left[ 
    \hat{\bf e}_\vartheta \frac{im P_l^m (\cos\vartheta)}{\sin\vartheta} 
    - \hat{\bf e}_\varphi \frac{{\rm d} P_l^m  (\cos\vartheta)}{{\rm d}\vartheta} 
    \right] \exp\left(im\varphi\right) \ ,
\nonumber
\\
{\bf Y}^{(o)}_L &= i \sqrt{ \frac{(l-m)!}{(l+m)!} } \sqrt{\frac{2l+1}{4\pi }} P_l^m (\cos\vartheta) \exp\left(im\varphi\right) \hat{\bf e}_r \ ,
\nonumber
\\
{\bf Y}^{(e)}_L &= i \sqrt{ \frac{(l-m)!}{(l+m)!} } \sqrt{\frac{2l+1}{4\pi l(l+1)}} \nonumber \\
& \times \left[ 
    \hat{\bf e}_\vartheta \frac{{\rm d} P_l^m (\cos\vartheta)}{{\rm d}\vartheta} 
    + \hat{\bf e}_\varphi \frac{im P_l^m (\cos\vartheta)}{\sin\vartheta} 
    \right] \exp\left(im\varphi\right) \ ,
\nonumber 
\end{align}
where $\hat{\bf e}_\varphi$, $\hat{\bf e}_\vartheta$, and $\hat{\bf e}_r$ are corresponding unit vectors, and $P^m_l(x)$ are the associated Legendre functions of the first kind~\cite{Abramowitz1973} of degree $l$ and order $m$:

\begin{equation}
P^m_l(x)=\frac{(-1)^m}{2^l l!} (1-x^2)^{m/2} \frac{{\rm d}^{l+m}}{{\rm d}x^{l+m}} (x^2-1)^l \ .
\nonumber 
\end{equation}

\subsection{Angular integration of multipole fields}
\label{sc:angint}
Recalling the fact that vector spherical harmonics are orthonormal, i.e.
\begin{equation}
    \oint \left| {\bf Y}^{(o,m,e)}_{L}({\bf r})\right|^2 {\rm d} \Omega = 1 \ , 
\nonumber
\end{equation}
one obtains for the multipole field ${\bf F}_{M L}$ of \eqref{fvmultip}:
\begin{equation}
\oint \left| {\bf F}_{M L}\right|^2 {\rm d}\Omega = \left|f_{Ml}(kr)\right|^2 \oint \left|{\bf Y}^{(m)}_L({\bf r})\right|^2 {\rm d}\Omega = \left|f_{Ml}(kr)\right|^2 \ .
\label{bdipideh2}
\end{equation}

After substituting from \cite[Eq. (10.1.22)]{Abramowitz1973} into \cite[Eq. (10.1.20)]{Abramowitz1973} in the recurrence relations for the spherical Bessel functions,

\begin{equation}
\frac{(y f_{l})'}{y} = f_{l}' + \frac{f_{l}}{y} = \frac{l+1}{2l+1} f_{l-1} - \frac{l}{2l+1} f_{l+1} \ ,
\label{bbff}
\end{equation}
and the radial derivative in the multipole field ${\bf F}_{EL}$ can be easily eliminated:

\begin{equation}
\begin{split}
{\bf F}_{EL}&(k_n,{\bf r}) =
\frac{\sqrt{l(l+1)}f_{EL} (k_n r)}{k_n r}\, {\bf Y}^{(o)}_L({\bf r}) \\
& + \left[ \frac{l+1}{2l+1}\, f_{E,l-1} (k_n r)
- \frac{l}{2l+1}\, f_{E,l+1}(k_n r)\right]\,{\bf Y}^{(e)}_L({\bf r}) \ .
\end{split}
\label{fela}
\end{equation}
On using the orthonormality of ${\bf Y}^{(o,m,e)}_{L}$, together with \cite[Eq. (10.1.19)]{Abramowitz1973} applied to the contribution resulting from $\sim {\bf Y}^{(o)}_L$ term, \eqref{fela} is transformed into:

\begin{equation}
\oint \left| {\bf F}_{E L}\right|^2 {\rm d}\Omega = \frac{l+1}{2l+1} |f_{E,l-1}|^2  + \frac{l}{2l+1} |f_{E,l+1}|^2 \ .
\label{bdipidee2}
\end{equation}

\eqref{bdipideh2} and \eqref{bdipidee2} are used to perform angular integration of \eqref{wdendef}, which yields in \eqref{e2} and \eqref{h2}.

\subsection{Lommel's integration formulas}
\label{sc:lomm}
Consider the defining {\em spherical} Bessel equation of given order $l$,

\begin{equation}
\dfrac{{\rm d}^2 \mathcal{F}_l(k r)}{{\rm d}r^2} + \frac{2}{r} \dfrac{{\rm d} \mathcal{F}_l(k r)}{{\rm d} r} + \left[k^2-\frac{l(l+1)}{r^2}\right] \mathcal{F}_l(k r) = 0 \ ,
\label{sbesse}
\end{equation}
where $r$ is real in our case, and $k$ is, in general, a complex parameter.

Much the same as in the case of the Lommel formula for the {\em cylindrical} Bessel equation~\cite[Chap. V]{Watson1995}, the corresponding Lommel formula for two arbitrary solutions $\mathcal{F}_l(\rho r)$ and $\mathcal{G}_l(\sigma r)$ (here, in general, $\sigma \neq \rho$ are arbitrary complex numbers) of the spherical Bessel equation \eqref{sbesse} with different parameter values follows straightforwardly from a fact that

\begin{equation}
\begin{split}
\frac{\rm d}{{\rm d} r} & \left[\mathcal{F}_l(\rho r) \frac{{\rm d} \mathcal{G}_l(\sigma r)}{{\rm d} r} - \frac{{\rm d} \mathcal{F}_l(\rho r)}{{\rm d} r} \mathcal{G}_l(\sigma r)\right] \\
& = 
\left[\mathcal{F}_l(\rho r) \dfrac{{\rm d}^2 \mathcal{G}_l(\sigma r)}{{\rm d}r^2} - \dfrac{{\rm d}^2 \mathcal{F}_l(\rho r)}{{\rm d}r^2} \mathcal{G}_l(\sigma r)\right]
\\
& =
-\frac{2}{r} \left[\mathcal{F}_l(\rho r) \frac{{\rm d} \mathcal{G}_l(\sigma r)}{{\rm d} r} - \frac{{\rm d} \mathcal{F}_l(\rho r)}{{\rm d} r} \mathcal{G}_l(\sigma r)\right] \\ 
& \qquad + (\rho^2-\sigma^2) \mathcal{F}_l(\rho r) \mathcal{G}_l(\sigma r) \ , \nonumber
\end{split}
\end{equation}
The latter expression can be recast as

\begin{equation}
\begin{split}
\frac{\rm d}{{\rm d} r} & \left[r^2 \left(\mathcal{F}_l(\rho r) \frac{{\rm d} \mathcal{G}_l(\sigma r)}{{\rm d} r} - \frac{{\rm d} \mathcal{F}_l(\rho r)}{{\rm d} r} \mathcal{G}_l(\sigma r)\right) \right] \\
& = r^2 \left(\rho^2-\sigma^2\right) \mathcal{F}_l(\rho r) \mathcal{G}_l(\sigma r) \ .
\nonumber
\end{split}
\end{equation}
On integrating both sides one immediately arrives at 

\begin{equation}
\begin{split}
(\rho^2-\sigma^2) &  \int^r \mathcal{F}_l(\rho r) \mathcal{G}_l(\sigma r) r^2 {\rm d} r
\\
= & r^2 \left[
\mathcal{F}_l(\rho r) \frac{{\rm d} \mathcal{G}_l(\sigma r)}{{\rm d} r} - \frac{{\rm d} \mathcal{F}_l(\rho r)}{{\rm d} r} \mathcal{G}_l(\sigma r)\right] + \mathcal{C} \ ,
\end{split}
\label{3dstrw}
\end{equation}
where $\mathcal{C}$ is an integration constant. 

In the limit $\rho\to\sigma$, the lhs of \eqref{3dstrw} goes to zero, whereas the rhs seems to be a nonzero function of $r$. After a closer inspection one finds that the square bracket on the rhs reduces to the Wronskian $W_r\{\mathcal{F}_l(\sigma r), \mathcal{G}_l(\sigma r)\}$. Any solution $\mathcal{F}_l$ and $\mathcal{G}_l$ of \eqref{sbesse} of order $l$ can be expressed as a linear combination of spherical Bessel functions $j_l$ and $n_l$ with, in general complex, coefficients [cf. \eqref{flcmb}]. Subsequently, the Wronskian $W_r\{\mathcal{F}_l, \mathcal{G}_l\}$ breaks down into a sum of terms proportional to $W_r\{j_l(\sigma r),j_l(\sigma r)\}$, $W_r\{n_l(\sigma r), n_l(\sigma r)\}$, and $W_r\{j_l(\sigma r), n_l(\sigma r)\}$. The first two factors are identically zero. The last one is proportional to $1/r^2$ \cite[Eq. (10.1.6)]{Abramowitz1973}, which cancels the factor $r^2$ in front of the square bracket in \eqref{3dstrw}. This means that the term with the square bracket on the rhs of \eqref{3dstrw} reduces to, in general, nonzero constant $\tilde{\mathcal{C}}$ the limit $\rho\to\sigma$. By taking the integration constant $\mathcal{C}$ in \eqref{3dstrw} to be just the opposite of the constant $\tilde{\mathcal{C}}$, both sides of \eqref{3dstrw} go to zero to the limit $\rho\to\sigma$ and the equality is preserved. The point of crucial importance is that for any specific pair of spherical Bessel functions $\mathcal{F}_l$ and $\mathcal{G}_l$ there is a unique constant $\mathcal{C}$ which ensures equality in \eqref{3dstrw} including the limit $\rho\to\sigma$. One can thus readily apply l'H\^opital rule to investigate the limit $\rho\to\sigma$ of the expression

\begin{equation}
\begin{split}
\int^r & \mathcal{F}_l(\rho r) \mathcal{G}_l(\sigma r) r^2 {\rm d} r =
\frac{r^2}{\rho^2-\sigma^2}
\\
& \times
\left[
\mathcal{F}_l(\rho r) \frac{{\rm d} \mathcal{G}_l(\sigma r)}{{\rm d} r} - \frac{{\rm d} \mathcal{F}_l(\rho r)}{{\rm d} r} \mathcal{G}_l(\sigma r) \right] + \frac{\mathcal{C}}{\rho^2-\sigma^2} \ .
\end{split}
\label{3dstrt}
\end{equation}
Since the constant $\mathcal{C}$ naturally disappears after applying l'H\^{o}pital's rule, or when performing definite integrals, it does not impact final expressions and will be omitted below.

If one wants to consider the special case when $\mathcal{G}_l=\mathcal{F}_l^*$, then, as the result of complex conjugation, also the argument of Bessel functions forming $\mathcal{F}_l$ becomes complex conjugated. An appropriate application of the Lommel's first integral (\ref{3dstrt}) to this special case is therefore

\begin{equation}
\begin{split}
   \mathcal{I}^{(1)}_{L} & = \int^r \mathcal{F}_l (\rho r) \mathcal{F}^*_l (\sigma r) r^2  {\rm d} r \\
    & = \dfrac{r^2}{\rho^2 - \sigma^{*2}} \left[ \mathcal{F}_l (\rho r) \dfrac{{\rm d}\mathcal{F}^*_l (\sigma r)}{{\rm d}r} - \dfrac{{\rm d}\mathcal{F}_l (\rho r)}{{\rm d}r} \mathcal{F}^*_l (\sigma r) \right],
\end{split}
\label{lomder}
\end{equation}
because $\mathcal{F}^*_l(\sigma r)$ satisfies the Bessel equation (\ref{sbesse}) with $k=\sigma^*$. After using the recurrence relation for Bessel functions \cite[Eq. (10.1.22)]{Abramowitz1973},

\begin{equation}
    \dfrac{{\rm d}\mathcal{F}_{l}(kr)}{{\rm d}r} = \dfrac{l}{r} \mathcal{F}_{l}(kr) - k \mathcal{F}_{l+1}(kr) \ , \nonumber
\end{equation}
one arrives at:

\begin{equation}
\mathcal{I}^{(1)}_{L} =\dfrac{r^2}{\rho^2 - \sigma^{*2}} \left[ \rho \mathcal{F}_{l+1} (\rho r) \mathcal{F}^*_{l} (\sigma r)
  - \sigma^* \mathcal{F}_{l} (\rho r) \mathcal{F}^*_{l+1} (\sigma r)
  \right] \ . \nonumber
\end{equation}
In the special case of $\sigma^*=\rho^*$:

\begin{equation}
\begin{split}
    \int^r & 
     \mathcal{F}_l (\rho r) \mathcal{F}^*_l (\rho r) r^2 {\rm d} r
     = \int^r
     \left| \mathcal{F}_l(\rho r) \right|^2 r^2 {\rm d}r \\
    & = \dfrac{r^2}{\rho^2 - \rho^{*2}} \left[\rho \mathcal{F}_{l+1} (\rho r) \mathcal{F}^*_{l} (\rho r) - \rho^* \mathcal{F}_{l} (\rho r) \mathcal{F}^*_{l+1} (\rho r) \right] \\
    & = \dfrac{r^3}{x^2 - x^{*2}} \left[ x \mathcal{F}_{l+1} (x) \mathcal{F}^*_{l} (x) - x^* \mathcal{F}_{l} (x) \mathcal{F}^*_{l+1} (x) \right] \ ,
\end{split}
\label{lom}
\end{equation}
where $x = \rho r$. The latter expression in square brackets corresponds to our \eqref{fpl}.

In the real limit $\rho^* \to \rho$, which corresponds to a lossless core or a lossless shell in the current work, an application of l'H\^{o}pital's rule and recurrence relations \cite[Eqs. (10.1.21-22)]{Abramowitz1973} for the Bessel functions in \eqref{lom} yield:

\begin{widetext}
\begin{equation}
\begin{split}
    \int^r \left| \mathcal{F}_l(\rho r) \right|^2 r^2 {\rm d}r 
    & = -\dfrac{r^2}{2\rho} \left[ \rho r \mathcal{F}_{l+1} (\rho r)  \dfrac{{\rm d}\mathcal{F}^*_l (\rho r)}{{\rm d}(\rho r)} - \rho r \mathcal{F}_{l} (\rho r) \dfrac{{\rm d}\mathcal{F}^*_{l+1} (\rho r)}{{\rm d}(\rho r)} - \mathcal{F}_{l} (\rho r) \mathcal{F}^{*}_{l+1} (\rho r) \right] \\
    & = \dfrac{r^3}{2x} \left[ x \left( \left| \mathcal{F}_{l}(x) \right|^2 + \left| \mathcal{F}_{l+1}(x) \right|^2 \right) - (2l+1){\rm Re}\left( \mathcal{F}_{l}(x) \mathcal{F}^*_{l+1}(x) \right) \right] \ . \nonumber
\end{split}
\end{equation}
\end{widetext}
\noindent
The latter expression in square brackets corresponds to our \eqref{ldsh}.


\end{document}